\documentclass[a4paper,11pt]{article}
\pdfoutput=1 % if your are submitting a pdflatex (i.e. if you have
\usepackage{jcappub} % for details on the use of the package, please
\usepackage{float}

\def\apj{ApJ}
\def\apjl{ApJL}

\def\araa{ARAA}
\def\jcap{J. Cosmol. Astropart. Phys.}

\def\prd{Phys. Rev. D}

\def\beq#1{\begin{equation}\label{#1}}
\def\eeq{\end{equation}}
\def\beqa#1{\begin{eqnarray}\label{#1}}
\def\eeqa{\end{eqnarray}}

\def\myfrac#1#2{\left(\frac{#1}{#2}\right)}

\def\mycomment#1{\relax}
\title{Electromagnetic radiation accompanying  
gravitational waves from black hole binaries}

\usepackage[T2A]{fontenc}
\usepackage[maccyr]{inputenc}
\usepackage[russian,english]{babel}

\def\be{\begin{eqnarray}}
\def\ee{\end{eqnarray}}

\author[a,b]{A. Dolgov,}
\author[c,d]{and K. Postnov}

\affiliation[a]{Novosibirsk State University,  630090 Novosibirsk, Russia}
\affiliation[b]{ITEP,  Bol. Cheremushkinsaya ul., 25, 117218 Moscow, Russia}
\affiliation[c]{Sternberg Astronomical Institute, Moscow M.V. Lomonosov State University,\\
 Universitetskij pr., 13,  119234 Moscow, Russia}
\affiliation[d]{National Research University Higher School of Economics, Myasnitskaya ul. 20, 101000 Moscow, Russia}

\emailAdd{dolgov@fe.infn.it}
\emailAdd{kpostnov@gmail.com}

\abstract{
The transformation of powerful gravitational waves, created by the coalescence of massive black hole 
binaries, into electromagnetic radiation in external magnetic fields is revisited. In contrast to the 
previous calculations of the similar effect, we study the realistic case of the gravitational radiation 
frequency below the plasma frequency of the surrounding medium. The gravitational waves propagating 
in the plasma constantly create electromagnetic radiation dragging it with them, despite the low frequency.
The plasma heating by the unattenuated electromagnetic wave may be significant 
in a hot rarefied plasma with strong magnetic field and can lead to 
a noticeable burst of electromagnetic radiation with higher frequency.
The graviton-to-photon conversion effect in plasma is discussed in the context of possible
electromagnetic counterparts of GW150914 and GW170104.
}

\keywords{gravitational waves, gravitational wave sources}
\arxivnumber{1706.05519}

\begin{document}
\maketitle
\flushbottom

\section{Introduction} \label{sec:intro}
 
Recent observations of gravitational waves (GW) by LIGO interferometers~\cite{LIGO-PRL,2016PhRvL.116x1103A,PhysRevLett.118.221101} from coalescing black hole (BH)
binaries compellingly confirmed two important predictions of Einstein's 
General Relativity -- the existence of gravitational waves and astrophysical BHs. The sky localization of GW sources by a network of GW interferometers 
is limited by long GW wavelengths to a few square degrees \cite{2016LRR....19....1A}, and therefore
for the GW astronomy it is important to search for the possible accompanying electromagnetic radiation from the 
source localization region.

In the observed LIGO's binary BH coalescences, the energy released in GW amounts to several solar masses.
Even if a tiny part of the powerful GW pulse transforms into photons, the effect could be extremely bright.
In this connection, the transformation of gravitational waves into electromagnetic ones in an external magnetic 
field~\cite{g-to-gamma1}-\cite{g-to-gamma7} might create an observable burst of electromagnetic (EM) radiation.
Contemporary calculations of this effect with an account for QED and plasma corrections are presented in refs.~\cite{RS,AD-DE}. However in these works the 
graviton-photon transformation was considered at rather high GW frequencies, $\omega/(2\pi)$, exceeding the plasma frequency 
of the surrounding medium, $\Omega/(2\pi)\simeq 10 \mathrm{kHz}\,\sqrt{n_e}$, where $n_e$ is electron number density.
The reason for that is evident: low-frequency electromagnetic waves with $\omega < \Omega$ 
do not propagate in plasma. 
The only known to us low-frequency example, with $\omega < \Omega$, was considered in paper \cite{2000ApJ...536..875M},
where the nonlinear generation of higher electromagnetic harmonics in plasma was studied. These 
harmonics can have sufficiently high frequency allowing them to propagate in the plasma and give rise
to potentially observable radio emission.

Importantly, in the case of the LIGO events (the coalescence of binary BHs with 20-30 solar masses),
the GW frequency $\omega/(2\pi)\sim 100-200$~Hz, and the condition $\omega <\Omega $
is indeed realized for astrophysically relevant plasma densities. Nevertheless, the energy transition from the gravitational radiation into electromagnetic energy is still possible.
In this paper we show that GWs propagating in
plasma with high $\Omega$ and non-zero magnetic field continuously transform a part of their energy into the non-propagating
plasma waves, which can noticeably heat up the plasma and in turn lead to a burst of electromagnetic radiation.

\section{GW-to-gamma transformation} \label{s-GW-gamma}

The transition of a plane gravitational wave, $\sim \exp (-i \omega t + i {\bf kx} )$, into an electromagnetic one in external
transverse magnetic field $B_T$ is described by the following equations (see e.g. ref.~\cite{RS} for the derivation):   
\begin{align}\label{system3}
(\omega^2 - k^2) %\partial_{\mathbf{x}}^2)
h_{j} (\mathbf{k}) &=\kappa k %\partial_{\mathbf{x}}
A_j(\mathbf{k})B_{T}\,,  \\  %\nonumber
(\omega^2 - k^2 -  m^2) % +i \omega \Gamma) %\partial_{\mathbf{x}}^2)
%A_j - {m^2} 
A_j (\mathbf{k})
& =\kappa k %\partial_{\mathbf{x}}
h_j(\mathbf{k})B_{T}\,,
\label{system4}
\end{align}
where $B_T$  is the external magnetic field component orthogonal to the graviton propagation, subindex $j$ defines the
polarization state of the graviton or photon, and $h_j$ is the canonically normalized field of the  gravitational wave, such that
the kinetic term in the Lagrangian has the form $ (\partial_\mu h_j )^2$. In other words, $h_j$ is related to the metric 
$g_{\mu\nu} = \eta_{\mu\nu}  +  \tilde h_{\mu\nu} $ according to the relation
\be
h_j = \tilde h_j /\kappa
\label{h-j}
\ee
where $\kappa^2 = 16 \pi /m_\mathrm{Pl}^2$, with $m_\mathrm{Pl} \approx 2\cdot 10^{19}$ GeV being the Planck mass.

The last term in eq.~(\ref{system4}) is the effective mass of a photon in the medium.  It includes, in particular, the
plasma frequency and the Heisenberg-Euler correction. Under the conditions of the problem,  $m$
is dominated by $\Omega$:
\be
m^2  =  \Omega^2 - \frac{2\alpha  C\omega^2 }{45 \pi}  \left( \frac{B}{B_c}\right)^2 \approx \Omega^2, 
\label{m2}
\ee
where $B_c = m_e^2/ e$ is the critical (Schwinger) magnetic field $\simeq 4.4 \times 10^{13}$~G, $e^2 = 4\pi\alpha = 4\pi/137$. and
$C$  is a numerical constant of  order unity. It depends upon the  relative directions of the magnetic field vector $\bf B$ and the wave
polarization. The plasma frequency is equal to:
\be 
\Omega ^2 = {n_e e^2}/{m_e},
\label{Omega-2}
\ee
where  $n_e$ is the density of electrons; the contribution of ions is neglected here.

As we have already mentioned, the frequency of the gravitational waves  registered by LIGO is small in comparison with the
plasma frequency of the interstellar medium. Therefore, in the ordinary interstellar medium the second (QED) term in eq.~ (\ref{m2}) can be neglected. However, in the
case of larger $\omega$ and/or  large magnetic fields the two terms in eq.~(\ref{m2}) may become comparable, and this would lead
to a strongly amplified resonance graviton-to-photon transition.

The eigenvalues of the wave vector of the system of equations (\ref{system3}), (\ref{system4}) are:
\be
k_1 =  \pm \omega \sqrt{1 + \zeta^2} ,\,\,\,  k_2 = \pm  i m \sqrt{(1 - \zeta^2)(1-\eta^2)}, 
\label{lambda2}
\ee
where
\begin{equation}
\zeta^2 = (\kappa B)^2 /m^2 \ll 1,  \,\,\,\, \eta^2 = \omega^2 / m^2\,. % \lambda = \bar \lambda /m .
\label{prmtrs}
\end{equation}
The following eigenfunctions correspond respectively to these eigenvalues: 
\be
A_1 = \eta\, \zeta h_1,\,\,\,\,
h_2 =  i\zeta  A_2 .
\label{Apm}
\ee
The first solution describes a graviton entering into the magnetic field and creating a little photons, while the second one, vice versa,
describes a photon which creates a little gravitons in the magnetic field. In the second case the wave vector $k$  is purely 
imaginary, which corresponds to damping of the electromagnetic wave in plasma when its frequency is below the plasma
frequency. In the first case the wave vector $k_1$ is real, and the electromagnetic wave is not attenuated and keeps on 
running together with the gravitational wave, despite its low frequency. The gravitational wave carries the electromagnetic 
companion and does not allow it to damp.

\section{Heating of plasma by the graviton to photon transition.}\label{s-gw-heat}

The interaction of an electromagnetic wave with medium is described by the dielectric permittivity $\epsilon$, which determines
the relation between the wave vector and frequency of the electromagnetic wave  $ k^2 = \epsilon \omega^2$. For
the first solution,  $k \approx \omega$  up to some small corrections of the order of  $\zeta^2$. But this is not all. We need to take 
into account the imaginary part of  $\epsilon $, which arises as a result of interactions of the electromagnetic wave with 
electrons in the medium. This imaginary part leads to transition of energy from the electromagnetic wave into the plasma. 
For transverse waves in collisonless plasma, this imaginary part is calculated e.g. in the book~\cite{LL-10}, the problem 2, eq. (5) after section 31:
\be
{\rm Im}\, \epsilon  = \sqrt{\frac{\pi}{2} }\,\frac{\Omega}{\omega k a_e} \approx \sqrt{\frac{\pi}{2}}\,\frac{\Omega}{\omega^2 a_e},
\label{Im-eps}
\ee
where $ a_e = \sqrt{T_e/ (e^2 n_e)}\simeq 743 \sqrt{(T_e/\mathrm{K})/(n_e/\mathrm{cm}^{-3})}$~cm is the Debye screening length for electrons and $T_e$ is their temperature. In the
last equation, the relation (\ref{lambda2}) $ k = k_1 = \omega$ is used.

In the collisionless plasma approximation, this lost energy goes  from the GW to the plasma and back. However, 
an account of the interaction of the 
electromagnetic wave with electrons in plasma leads to the heating of plasma by energy of photons
created by the
gravitational wave. If the heating happens to be non-negligible, then an excessive EM radiation from the 
heated plasma may be registered.

For the interstellar medium with electron density $n_e = 0.1\, {\rm cm}^{-3}$ and temperature $T_e = 1 $ eV,
the Debye length is approximately equal to 
$a_e \approx 10^3\,\rm{cm} =3\cdot 10^{-8}$ seconds,  the plasma frequency is about 
$ \Omega \approx 3\cdot 10^4 \,{\rm sec}^{-1}$, while the frequency of the first registered LIGO event is 
$\omega \approx 2000$~rad~s$^{-1}$. Correspondingly, $\Omega a_ e \approx 10^{-3}$ and thus
$\omega^2 {\rm Im}\,\epsilon \sim \Omega / a_e$ is much larger than $\Omega^2$. Therefore, the  amplitude of
the electromagnetic wave, carried by the gravitational wave  is given by the equation:
\be
A_j \approx  \frac{\omega a_e \kappa B}{\Omega} \, h_j
% \frac{\omega a_e}{\Omega^2 }\,   h_j  = \left( \frac{\omega}{\Omega}\right)^3 \,a_e \kappa B.
\label{A-of-h}
\ee
So the energy flux of photons absorbed by the plasma makes the following fraction of the energy flux of the
parent gravitational wave:
\be
K \equiv \frac{\rho_\gamma }{\rho_{GW}} =\left( \frac{\omega a_e \kappa B}{\Omega} \right)^2
\approx 10^{-46}\myfrac{\omega}{\Omega}^2\myfrac{a_e}{\mathrm{1\,cm}}^2 \myfrac{B}{1\,\mathrm{G}}^2\,.
\label{fraction}
\ee 

According to \cite{LIGO-PRL,PhysRevLett.118.221101}, the total GW energy emitted in LIGO binary black hole coalescences is about $3 M_\odot$ 
during approximately 0.01 seconds. So the GW energy flux at the distance $R$ from the source is  
\be
F_{GW} \approx 100 M_\odot / (4 \pi R^2)\, {\rm per}\,\,\, {\rm second }.
\label{rho-GW}
\ee

The frequency of these gravitons (and the produced photons) is a few hundred Hz, which is by far smaller than 
the temperature of the
interstellar or intergalactic media and even smaller than the temperature of CMB (in natural units $c=\hbar=k_B=1$). Thus, even if plasma absorbs
a lot of energy, the direct heating of plasma by the EM wave would not be efficient. However, this is not all the truth because
the electrons in the plasma can be accelerated by the electric field of the running electromagnetic wave and acquire a very
large energy. Indeed, the electrons in the electric field of the wave are accelerated according to the equation:
\be
m_e \ddot x_e = e E = e E_0 \cos (\omega t)
\label{ddot-x}
\ee
and acquire the velocity 
\be
V_e \sim \ddot x_e /\omega \sim e E_0 / (m_e \omega),
\label{V-e}
\ee
where $ \omega $ is of the order of the frequency of the incoming gravitational wave. So the electrons could gain the energy:
\be
{\cal E}_e = \frac{m_e V_e^2 }{2} \sim \frac{e^2 E_0^2}{ m_e \omega^2} .
\label{E-e}
\ee
This result is true if the electron collision time due to Compton (Thomson) or Coulomb scattering is much longer than the
inverse frequency of the wave. This condition is normally fulfilled for the interstellar or intergalactic plasma.

If we take the distance $R$ equal to the gravitational radius of a black hole with mass 30 $M_\odot$, 
i.e. $R = r_g = 10^7$ cm, then the electrons  would be accelerated up to the energy  
$ {\cal E}_e = 4 {\rm eV} (B/{\rm G})^2 $, becoming relativistic for rather mild fields $ B \gtrsim 10^3$~G.
In such a plasma, $e^+e^-$ pairs must be created. Their presence would change the values of the plasma frequency and 
of the Debye length, but qualitatively the picture would remain essentially the same.

The above estimate is obtained under the assumption of a homogeneous external magnetic field, i.e. for the case
where the GW wavelength $\lambda$ is much smaller than the scale of the field homogeneity, $l_B$. In the opposite limit, the GW-EM conversion effect
would be suppressed by the factor $l_B/\lambda$.

\section{Possible observable effects }\label{s-obs-eff}

If such a gravitational wave falls on a magnetar with superstrong  magnetic field of about $10^{15}$~G, the produced burst of electromagnetic 
radiation would be significant for distances between the magnetar and the coalescing black holes of order of one astronomical unit,
which is very small by the astrophysical scales.
 
Much more plausible could be the burst of electromagnetic radiation if the BH binary itself is surrounded by a medium 
with sufficiently strong magnetic field. Such a field might be created in analogy with the  Biermann battery~\cite{biermann} 
induced by the 
rotating space-time around the binary due to the different mobility of protons and electrons in the surrounding
bath of electromagnetic radiation (see, e.g., \cite{1972JETP...34..233M} discussing the 
battery effect in application to the origin of the seed magnetic field in rotating protogalaxies immersed in an isotropic CMB photon field). In the battery mechanism on thermal electrons, the amplitude of the magnetic field 
is directly proportional to the energy density of photons $\epsilon_\gamma$ 
\beq{}
\frac{e}{m_ec}B\sim \frac{4}{3}c\sigma_T\frac{\epsilon_\gamma}{m_ec^2}(2\omega_{LT}\Delta t)+B_0\,,
\eeq
where $B_0$ is the initial magnetic field in the plasma, $\sigma_T$ is the Thomson cross-section, 
$\omega_{LT}$ is the Lense-Thirring frequency, $\Delta t$ is the field growth time (in this formula the velocity of light $c$ is recovered). Clearly, 
in the problem under consideration $\omega_{LT}\Delta t\sim 1$, and for typical interstellar (or CMB) photon 
energy density $\epsilon_{CMB}\sim 1$~eV~cm$^{-3}$ the produced field is insignificant:
\beq{e:battery}
B\sim 2\times 10^{-27}[\mathrm{G}]\myfrac{\epsilon_\gamma}{\epsilon_{CMB}}+B_0\,.
\eeq

The linear dependence of the battery effect on photon energy density suggests the most favorable sites 
for GW-EM conversion in the vicinity of strong radiation sources or in a hot plasma. For example, 
if 
a BH-BH merging happens close to an active galactic nucleus with typical luminosity of $\sim 10^{44}$~erg/s, $\epsilon_\gamma\sim L/(R^2c)\sim 2\times 10^8 (L/10^{44}\mathrm{erg\,s}^{-1})(R/1 \mathrm{pc})^2$~eV~cm$^{-3}$. 
In a hot rarefied cosmic plasma with $T\sim 10$~keV (e.g., in galaxy cluster centers) $\epsilon_\gamma\approx \tau (T/T_{CMB})^4\epsilon_{CMB}$, where $\tau$ is the characteristic optical depth. In our problem, 
$\tau\sim \lambda n_e \sigma_T$, so that $\epsilon_\gamma/\epsilon_{CMB}\sim (H_0/\omega)(T/T_{CMB})^4
\sim 10^{15}$. Both estimates, however, show that in real astrophysical conditions 
the battery effect could hardly lead to the fields which are interesting for the efficient GW-EM conversion. 

%The energy density of the thermal photon field scales as $\sim T^4$, 
%therefore if the plasma around the black hole binary is heated up to a high temperature, the battery effect 
%could be enhanced. For example, in the $T=10$ keV intergalactic gas with optical depth
%for Thomson scattering $\tau\sim 0.1$, $\epsilon_\gamma\approx \tau (T/T_{CMB})^4\epsilon_{CMB}\sim 10^{29}\epsilon_{CMB}$, leading to $B\sim 200$~G. This simplified estimate shows that should BH-BH coalescence 
%happens in an intensive X-ray or
%gamma-ray background, a strong
%electromagnetic  burst produced by the gravitational wave still can be expected from the coalescence. 

The rate of binary BH coalescences derived from LIGO observations is $\sim 12$ per cubic Mpc per year \cite{PhysRevLett.118.221101}. This rate 
is orders of magnitude smaller than the rate of bright short electromagnetic transient phenomena, such as short gamma-ray bursts (SGRB, about one per a few days for distances up to a few Gpc) or enigmatic fast radio bursts (FRB, a few thousand per day for distances up to hundreds Mpc). The mean energy emitted by an FRB is $\Delta E_{FRB}\simeq 10^{38}$~ergs \cite{2017RAA....17....6L}. Even to intercept this energy from a BH-BH merging, the magnetar should occur within a planetary system distance of a few astronomical units, which is quite improbable. The SGRB phenomenon, in turn, with a typical EM energy release of $10^{49}$~erg, is likely to be unrelated to BH-BH mergings \cite{2014ARA&A..52...43B}. 

However, the close time occurrence of a faint Fermi GBM event $\sim 0.5$~s after GW150914 \cite{2016ApJ...826L...6C}
and the recent AGILE observation of the possible short MeV burst preceding the final coalescence of
GW170104 by $\sim 0.5$~s \cite{2017arXiv170600029V}, if real, may suggest an efficient EM energy release during BH-BH coalescence.
A possible scenario is proposed in \cite{2017arXiv170604211D} assuming fragmentation of the 
collapsing core of a rapidly rotating star. The GW-EM conversion mechanism discussed in the present paper 
suggests that for the plausible conditions of dense plasma disk surrounding a coalescing compact binary
(see, e.g., the discussion in \cite{2017arXiv170504695S}) the magnetic field can reach
the equilibrium values $B^2/(8\pi)=\epsilon_B\rho$ and be as high as $10^{15}$~G for $\epsilon_B\sim 1$. 
Under the extreme conditions discussed in \cite{2017arXiv170504695S}, the  electron temperature
is limited to the pair creation values $T=\epsilon_T m_e$ with $\epsilon_T\sim 1$, and the Debye length is thus $a_e^2\sim \epsilon_T(m_em_p)/(4\pi\alpha\rho)=\epsilon_T/\Omega^2$.
Plugging these extreme values 
into equation (\ref{fraction}), we find the limit GW-EM efficiency conversion coefficient 
\beq{Klim1}
K\leq \epsilon_B\epsilon_T \frac{32\pi}{\alpha} \myfrac{\omega}{\Omega}^2\myfrac{m_e}{m_p}\myfrac{m_p}{m_\mathrm{Pl}}^2\sim 10^{-37}\epsilon_B\epsilon_T\ll 1\,.
\eeq

The small factor $(\omega/\Omega)^2$ further diminishes this estimate, taking into account
the GW frequency at the peak GW emission being around twice the Keplerian orbital value prior to the merging 
$\omega^2\sim (1/2)(m_\mathrm{Pl}/M)^2m_\mathrm{Pl}^2$, where $M$ is the total mass of the coalescing binary:
\beq{oO}
\myfrac{\omega}{\Omega}^2\simeq \frac{1}{2}\myfrac{m_\mathrm{Pl}}{M}^2\frac{m_\mathrm{Pl}^2(m_pm_e)}{4\pi\alpha\rho}
\approx \myfrac{3M_\odot}{M}^2\myfrac{\rho}{10^{-24}\mathrm{g\,cm}^3}^{-1}\,.
\eeq
For GW-EM conversion regime in plasma considered here to be
applicable, the GW to plasma frequency ratio $\omega/\Omega$ ratio should be smaller than one, suggesting
the total BH-BH binary lower mass limit $M\gtrsim 3 M_\odot/\sqrt{\rho/10^{-24}\mathrm{g\, cm}^{-3}}$,
which obviously holds for solar-mass BH-BH binaries in any realistic astrophysical conditions. 

Thus, we conclude that the GW-EM conversion effect in plasma considered here can hardly provide
significant plasma heating in LIGO BH-BH coalescences and explain 
the observed EM power in gamma-ray events possibly associated with
GW150914 and GW170104 mentioned above. However, the effect (\ref{fraction}) may turn out to be substantial
in hot rarefied plasma where the strong external 
magnetic field is dynamically unrelated to the plasma density (e.g. near the magnetars). 

%Therefore we conclude that 
%the GW-EM conversion effect considered in the present paper 
%can give rise to only very faint EM transients associated with the stellar BH-BH mergings.

%It is tempting to identify the suggested mechanism of the gravitational wave transition into electromagnetic radiation 
%with the driver of the fast radio bursts (FRB)~\cite{FRB}. These bursts of radio waves have huge power {\bf which?} and 
%according to the observational estimates flare up in the sky with the frequency ??? per year. Though the number of BH
%binaries, according to our model~\cite{bdpp}, can be sufficiently high, but the probability of finding a magnetar near the 
%binary is quite low. Moreover, this mechanism hardly can explain the observed FRG repeater. 

%Last but not least, especially from the point of view of the Earth inhabitants, energetic electromagnetic radiation 
%created by the gravitational waves from the coalescing BH binaries in the Galaxy in our neighborhood might have a dangerous
%impact on our planet.

\acknowledgments

%The authors thank Prof. A.V. Zasov for useful discussions. 
A.D. acknowledges the Russian Science Foundation Grant 16-12-10037 (the calculation of
the GW-EM conversion effect)
K.P. acknowledges support from RSF grant 16-12-10519 (the analysis of
observational consequences of the effect).

%\bibliography{puzzlesBH}

%\bibliographystyle{JHEP}
%\begin{thebibliography}

\providecommand{\href}[2]{#2}\begingroup\raggedright\endgroup

\end{document}